**Dynamic patterns of overexploitation in fisheries.**

Ilaria Perissi (*,1), Ugo Bardi (1, 2), Toufic El Asmar (3), Alessandro Lavacchi (4)

1. Consorzio Interuniversitario Nazionale per la Scienza e la Tecnologia dei Materiali (INSTM), Firenze, Italy.

2. Dipartimento di Scienze della Terra, Università degli Studi di Firenze, Italy.

3. Food and Agriculture Organization, FAO, Roma, Italy

4. Consiglio Nazionale delle Ricerche, CNR-ICCOM, Firenze, Italy.

*Correspondence to: ilariaperissi@gmail.com

**Abstract**

Understanding overfishing and regulating fishing quotas is a major global challenge for the 21st Century both in terms of providing food for humankind and to preserve the oceans' ecosystems. However, fishing is a complex economic activity, affected not just by overfishing but also by such factors as pollution, technology, financial factors and more. For this reason, it is often difficult to state with complete certainty that overfishing is the cause of the decline of a fishery. In this study, we developed a simple dynamic model based on the earlier, well-known Lotka-Volterra model or Prey-Predator model. To describe exploitation patterns, we assume that the fish stock and the fishing industry are coupled stock variables in the model and they dynamically affect each other, with the fishing yield proportional to both the fishing capital and the fish stock. The model is based on the concept that the fishing industry acts as the "predator" of the resource and that its growth and subsequent decline is directly related to the abundance of the fish stock. If the model can be fit historical data relative to specific fisheries, then it is a strong indication that the fishing industry is heavily affected by the magnitude of the fish stock and that, in particular, the decline of the yield and the decline of the stock are linked to each other. The model doesn't pretend to be a general description of the fishing industry in all its varied forms; however, the data reported here show that the model can indeed qualitatively describe several historical case of the collapse of fisheries. The model can also be used as a qualitative guide to understand the behavior of several other fisheries. These result indicate that one of the main factors causing the present crisis of the world's fisheries is the overexploitation of the fish stocks.

**Introduction.**

Many of the world's fisheries are showing a decline in the fishing yield, a phenomenon that's often interpreted in terms of the overexploitation of the fish stock. That is, the fishing industry appears to be consistently exploiting the resource to a level higher than the carrying capacity of the system [1,2]. In economics, the depletion of fisheries was explored first in some early studies by Scott Gordon [3] and Milner Schaefer [4]. In the general field of resource overexploitation, an important influence was the work by Garrett Hardin[5], known under the name of "The Tragedy of the Commons." Hardin's model was only qualitative, but it established the patterns of overexploitation of any resource that's produced at a rate faster than it can reform [6]. In more recent times, fisheries have been extensively modeled, normally with considerably complex models (see, e.g. Thomas Schoener [7]) and as reviewed, for instance, by Worm et al.[8].

Of course, in modeling fisheries several factors are taken into consideration in addition to overfishing including climate change[9]. Nevertheless, the mechanism of overexploitation may play a central role in the collapse of at least some fisheries and this could be demonstrated, in principle, by verifying that the yield of the industry is directly proportional to the fish stock. In order to study this issue, we developed a simple system dynamics model based on some elements of the well-known Lotka-Volterra or "prey-predator" model. Alfred Lotka[10] and, independently, Vito Volterra[11,12] were the first to use differential equations in order to describe the dynamics of the predator/prey interaction in biological systems and the structure of the model can be seen as an ancestor of modern "system dynamics" [13]. George Gause [14] was probably the first to seek for experimental validation of the Lotka-Volterra (LV) model and he found it, but only for very simple biological systems in the form of two species of yeasts in laboratory conditions. In general, the behavior of real biological systems turned out to be too complex to be captured by the simple LV model[15]. However, the model had been originally proposed by Vito Volterra, as describing the behavior of human fisheries, rather than biological systems, even thought, at that time, suitable mathematical tools to fit experimental data were not available[16]. On the basis of this early idea by Volterra, we developed a simplified version of the model and used modern tools of data fitting. In this simplified version, we assumed that the fish stock behaves as a non-renewable resource when it is exploited so fast that the reproduction rate becomes a negligible parameter in the system. Some data on this approach had been reported in an earlier paper[17] in regard of the mining industry and it can be shown that this dynamic model is equivalent to the well-known "Hubbert model", well known in studies of the exploitation of mineral resources. Owing to the general validity of the model, we adopt here a notation in which a 'Resource' with the variable 'R' is exploited, while the 'Capital', with the variable 'C', is the effort necessary to extract it from the stock. Even though the model we developed does not claim to be able to describe all the complex ecosystem and economics interactions that occur in a fishery, we can report several cases in which it is possible to use the model to describe the historical production patterns of fisheries. We believe that this approach can play an important role in helping people to understand the basic mechanisms of fishery management, and in particular of overexploitation.

**Methods.**

The model utilized here is based on the following couple of differential equations

$R' = -k_1 C R$

$C' = k_2 C R - k_3 C$

where "R" stands for the resource stock while "C" stands for the capital stock. It is easy to recognize that this model is similar to the well-known Lotka-Volterra, predator-prey model, except in the fact that it lacks the term for the reproduction of the prey (named the resource). Some tests were performed by a more complex form of the model where the reproduction term was added, but we found that it produced no significant improvement in the fitting of the historical cases reported in the following. The three constants of the model describe how efficiently fish is caught ($k_1$), how efficiently the fish stock is transformed into capital ($k_2$) and how rapidly capital is dissipated ($k_3$). The dimensions of the constants depend on the units used for the capital and resource stocks.

In figure 1, qualitative graphical solutions for some of the parameters of the model are reported. The simplified LV model was implemented using MATLAB® computing language and the associated Simulink toolbox.

All the fittings reported here were generated using the unconstrained nonlinear optimization method based the Nelder-Mead algorithm ("fminsearch", Matlab). The objective function is the sum of the square of residuals (SSE, sum of squared errors of prediction) represented by the deviations of the LV predicted data from actual empirical values of data. The fitting procedure was found to be very sensitive on the initial value of the $k_1$, $k_2$ and $k_3$, as well as to the initial values of the stocks (Ro and Co). The initial guess of the parameters was provided by Simulink Design Optimization toolbox. Because data of R' and data of C can be very different, to obtain a good representation of the model, we normalized all data series. The Goodness of fit (GOF) is provided calculating the Normalized Means Square Errors (NMSE) function. NMSE measure the discrepancy between real and the model estimated value. The NMSE value is calculated by the Matlab Curve Fitting toolbox: NMSE equal to 1 represents the perfect fit, NMSE equal to zero means that real data are no better than a straight line at matching the model.

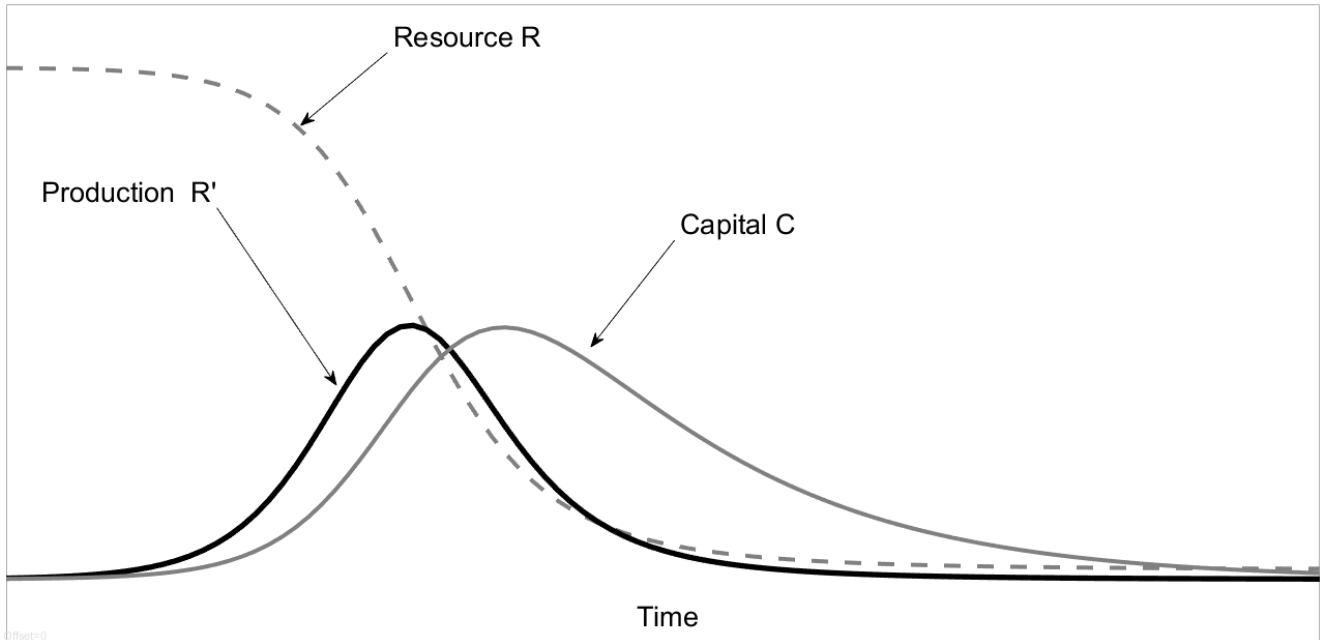

**Figure 1.** Qualitative solutions of the model. Neglecting the rate of prey reproduction imposing $k_0=0$, the prey-predator dynamic experiences a single oscillation, showing a definitive depletion of the prey stock and the subsequent collapse of predators' stock.

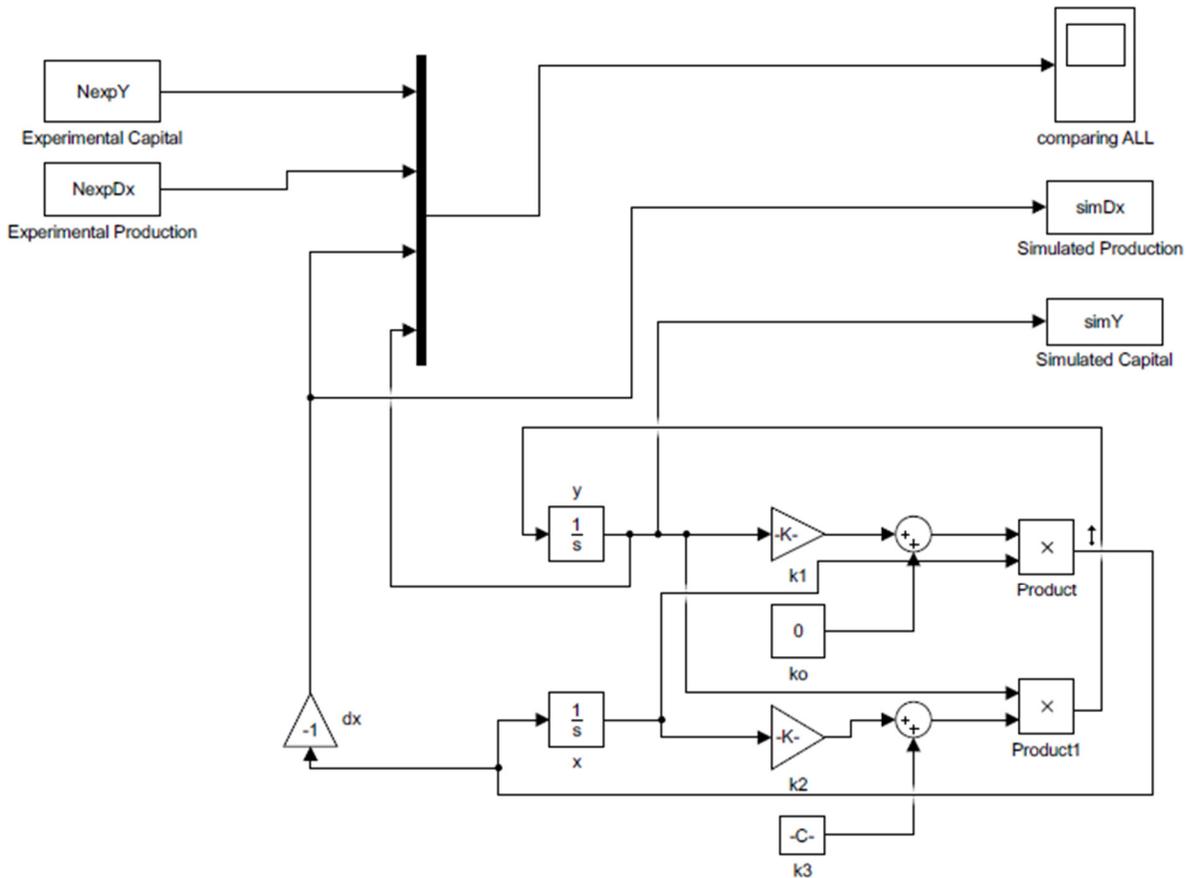

**Figure 2**. Simulink blocks' diagram of the Lotka Volterra model. Simulink provides modeling and simulation environment to edit the Lotka Volterra equations in a graphical form. NexpDx and NexpY are, respectively, the historical fish production, R', and the capital stocks. C; SimDx and SimY are the simulated production and capital: they correspond to the fitting curves.

**Results.**

Most of the systems examined here are "non managed" fisheries, named fisheries for which neither catch nor fishing effort is regulated or managed in any manner: in other words, every fisherman or fishing firm try to optimize their revenues by maximizing their catch. Historical data of these fisheries often show "bell shaped" trajectories of the fish landing data. Fish landing data represent the fish production, that is the quantity of fish that can be extracted by the fish stock, or in other words, the 'stock outflow', that in the model is indicated by R'.

Of course, the landing data, alone, are not sufficient to prove that prey-predator dynamic occurs, because we need to identify, also a 'predator stock'. We assumed the predator stock being the "capital" engaged in the fishing activity: in a free market economy more capital is usually invested to obtain more production. Whereas data on fishery production (landing) are often available, a direct measurement of the capital investments in fishing are not. For this purpose, therefore, we used "proxy" data, assumed to be proportional to the fishing effort, such as, for instance, the labor force employed in the fishery industry in number of persons or in terms of their total salary, the number of fishing vessels or the tonnage of the fleet. In recent statistical surveys, also the data on the investment in currency or as measured by economic indexes of the fishing sector are available. In fact, while in the past, in particular before the second world war, the number of boats or the boat tonnage really were proportional to the power in fishing, nowadays, the effort in fishing is more accurately represented by the required money to achieve technological improvement in fishing, modern equipment and boats, that not necessary implies a more copious fleet or number of fishermen.

Principally, the data we fitted were originated from "non managed" historical fisheries, but we observed the same dynamic also for some managed fisheries, probably indicating the shortcomings of the quota system. Note that the approach taken here always implies the fitting of at least two, independently measured, parameters of the system, R' and C. This "curve fitting" requires a complex optimization procedure of highly nonlinear equations that must be solved numerically at each iteration cycle. Not just any system can be described in this way. When the model fit lead to a good agreement with the historical data, that's a quite good indication that the real world stocks, fish and capital invested in fishing, interact each other in the way described by the model, a "predator/prey" relationship.

Following we show the results that tread the history of fisheries overexploitation since 19th century until the recent years. We tested the simplified LV model on different sets of free access data, that are not easily available both for capital and production on the same lapses of time.

Moreover, even thought, all possible fisheries data are not explored within this work, we think that our study is sufficiently robust to show that the overexploitation is an ongoing threat and that this is mainly due to a prey-predator dynamic.

**American whaling on 19th century.** The first documented evidence of the overexploitation of a marine resource goes back approximately two centuries ago, when the American whaling industry registered an intensive development due to the growing demand of whale derived products as the precious sperm oil, used to produce candles and lamp oil, and the whale bones and teeth that were used to produce such objects as flatware, corsets, toys, so that they are considered as 'the plastics of the $19^{th}$ century.' Here we examined the data reported in Alexander Starbuck's book about the history of the American whale fisheries from 1807 to 1876[18]. The sum of the production of oil from 'right' and 'sperm' whales, the two main species hunted, is assumed to be proportional to the predation activity and it is expressed in Gallons of oil.

The proxy for the predator stock is assumed to be the total tonnage of the whaling fleet. A graphical representation of the fitting is shown in figure 3. These data could be well fitted by the simplified LV model, as also reported in an earlier study [19].

The definitive collapse of the American whaling industry occurred at the beginning of $20^{th}$ century, as reported by Walter Tower [20] and more recently by Granville Mawer [21]

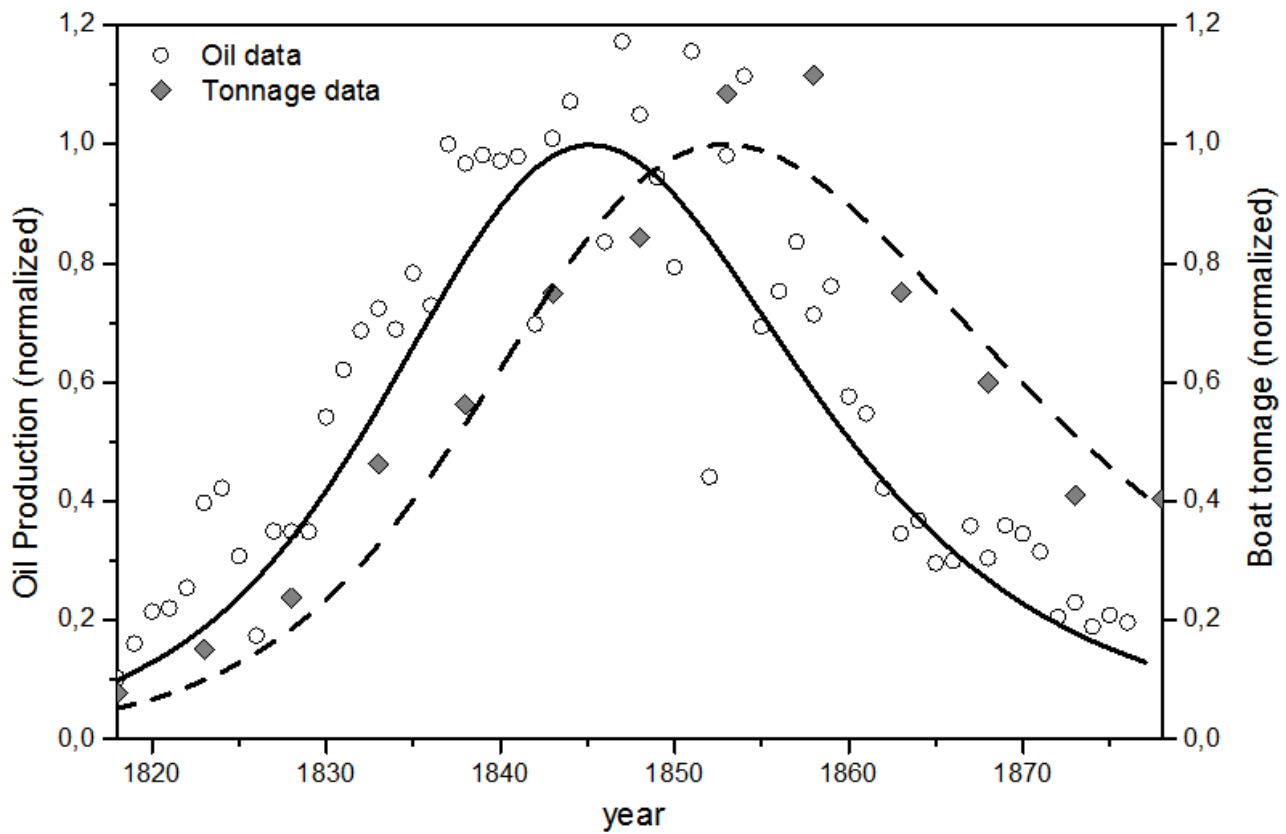

**Figure 3. Lotka Volterra modeling of American Whaling on 19<sup>th</sup> century**. Sperm Oil (production-prey) and the Tonnage capacity of fishing boats (capital-predator) from 1818 to 1878. Normalizing factors: oil 1.16 10$^5$ gallons, Boat tonnage 9.72 10$^4$ . Data Source: Starbuck, A. (1989). *History of the American Whale Fishery*. Castle. The GF for the fitted data: NMSE Oil fit: 0.76; NMSE Tonnage fit: 0.85.

**The collapse of the California Sardine Fishery**. The Pacific sardine fishery began to operate in central California in the late 1800s and grew in response to the increasing demand for food during World War I [22]. From the mid-1930s to the mid-1940s this fishery was the largest in the Northern and Southern West Pacific coasts, with peak landings of over 790,000 in 1936-1937 and average landings over 600,000 tons per season. The fishery began to collapse a few years afterward and catches declined over the next two decades to less than 100 tons per year in the 1970s[23]. Then, a moratorium was applied to the sardine fishing from 1974 to 1986[24]. In figure 4, we report the Californian sardine catch as production in comparison to the fishing fleet number of boats as capital, from 1933 up to 1957 and the related fitting obtained by LV modelling.

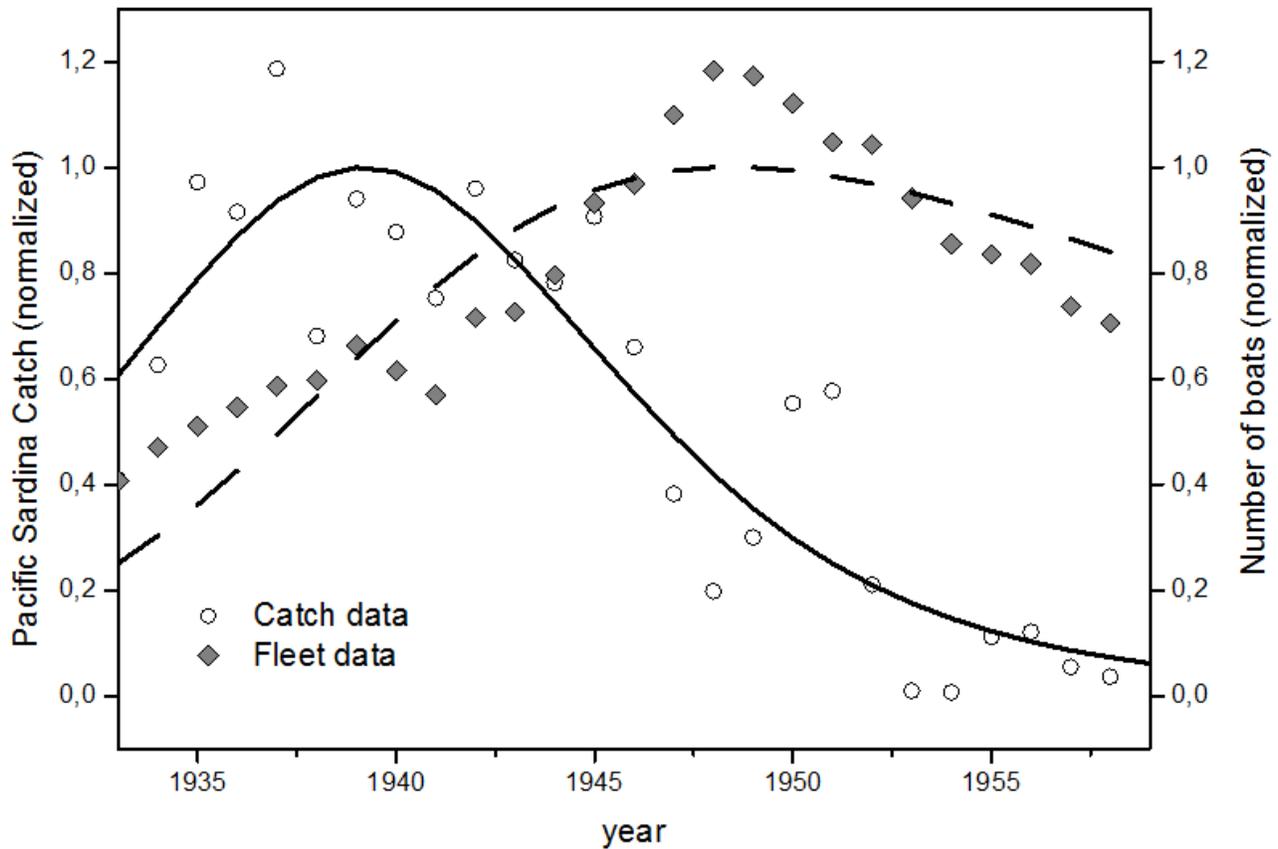

**Figure 4. Lotka Volterra modeling of Californian Sardines Fishery**. Catch (production-prey) and the number of fishing boats (capital-predator) from 1933 to 1959. Normalizing factors: catch 6.12 $10^5$ Tons, number of boats 5.21 $10^3$. Data Source: Fish Bulletin's, Fish Catch Statistics, California explores the Ocean website (https://library.ucsd.edu/ceo/). The GOF for the fitted data: NMSE Catch fit: 0.77; NMSE Fleet fit: 0.75.

The model shows how the capital under the form of number of boats reached the peak ten years later than production peak, leading to the fishery collapse. Again, we see that the Lotka-Volterra model provides a reasonably good fitting of the experimental data, indicating that the sardine stock was overexploited by the fishing industry.

**Japan Fishery industry.** The prey-predator dynamic can describe not only the exploitation of a single species; it is also able to illustrate the exploitation tendency of a whole fishing sector of a country. Here we report the evolution of the Japanese total fish catch and the national Disbursement of Fishery, from 1962 to 2002. The catch data are expressed as the quantity, in weight, of fish. The disbursement is expressed in currency and it includes expenses for the fisheries' in terms of wages, fuel, fishing boats capital and replacement and equipment, thus, it is a direct measurement of the capital investment in the sector. As shown in the figure 5, the historical data show a decline in the fishery yield starting with approximately 1980. The fitted peak of the capital expenditure occurs later, although the oscillations of the data indicate an earlier peak.

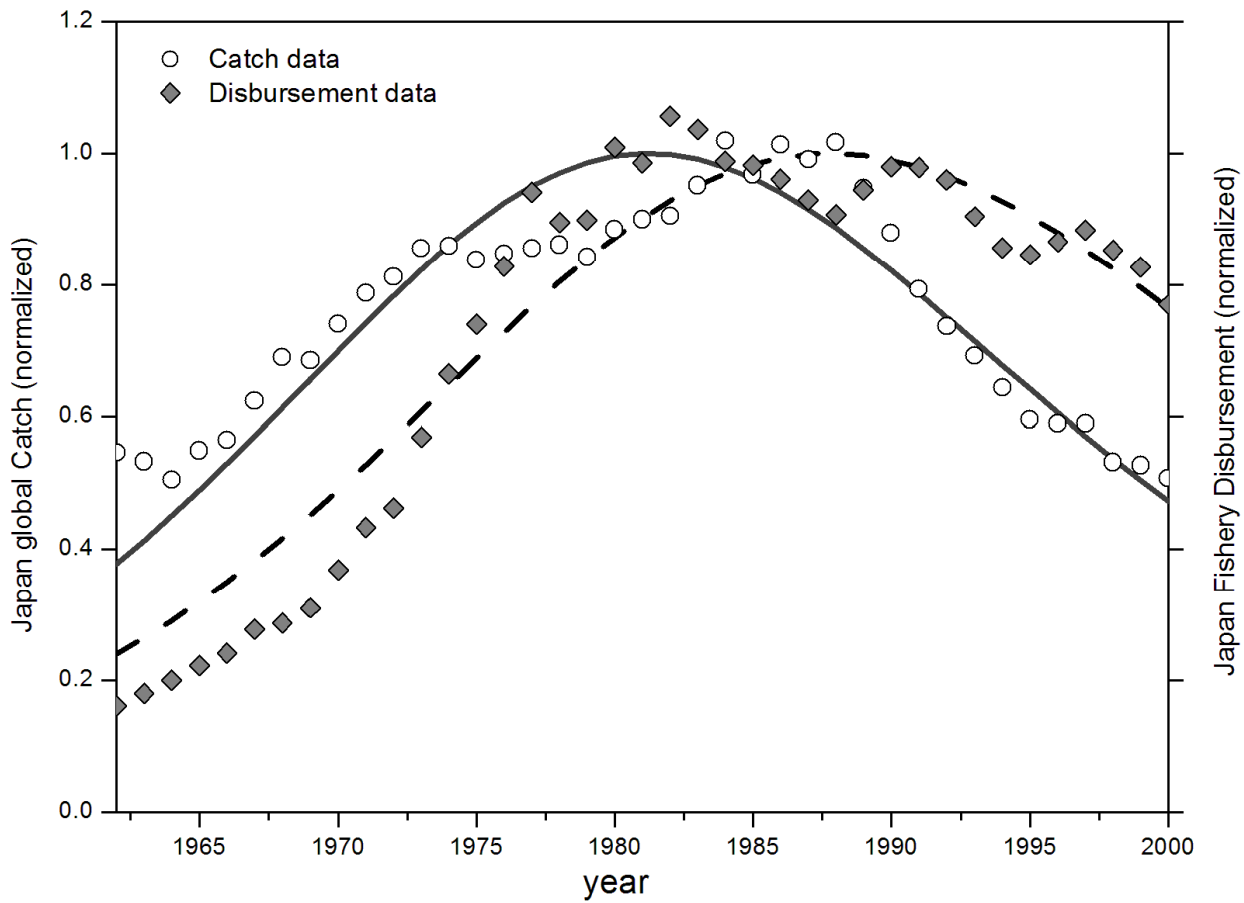

**Figure 5. Lotka Volterra modeling of Japanese Fishery Industry**. Total Catch (production-prey) and the Disbursement of Fishery (capital-predator) from 1962 to 2000. Normalizing factors: catch $1.26 \cdot 10^7$ Tons, disbursement $1.35 \cdot 10^8$ Yen. Data Source: Statistics Bureau, Ministry of Internal Affairs and Communications (http://www.stat.go.jp/). The GOF for the fitted

data: NMSE Catch fit: 0.81; NMSE Disbursement fit: 0.92.

Recent data from the Statistical Handbook of Japan 2015[25] show that, since 2000, the Japanese fish production trend is still declining. The value of the catch is decreasing with a rate of 25% from 2000 to 2014. For the same period, the Statistical Handbook of Japan 2015 also reports the number of Enterprises and the number of workers engaged in the Fishery sector. The values of such entities, even though they are not expressed in currency, can be reasonable assumed proportional to the capital effort invested in the sector. The data show that the trend, for both, is declining: in particular, from 2000 to 2014, the number of enterprises is reduced by 39%, while the number of workers is reduced by 33%.

**Iceland Fishery industry.** Iceland is another land whose economy was historically based on fishing, in particular cod and herrings, which still today represent the largest stocks. In the 1970s, the Icelandic fishing zone has been the theatre of the so-called 'cod wars' which saw the Icelandic fleet engaged in competition against foreign fleets. By the early 1970 the most valuable cod fish stocks had declined to very low levels of catch as a result of overfishing. In the same period, the north Atlantic Icelandic spring and summer spawning herring stock collapsed. The spring stock has not recovered yet, although the summer stock did [26,27]. The herring collapse provoked, as show in figure 6, the drastic fall of the number of Herring Salting industries few years later.

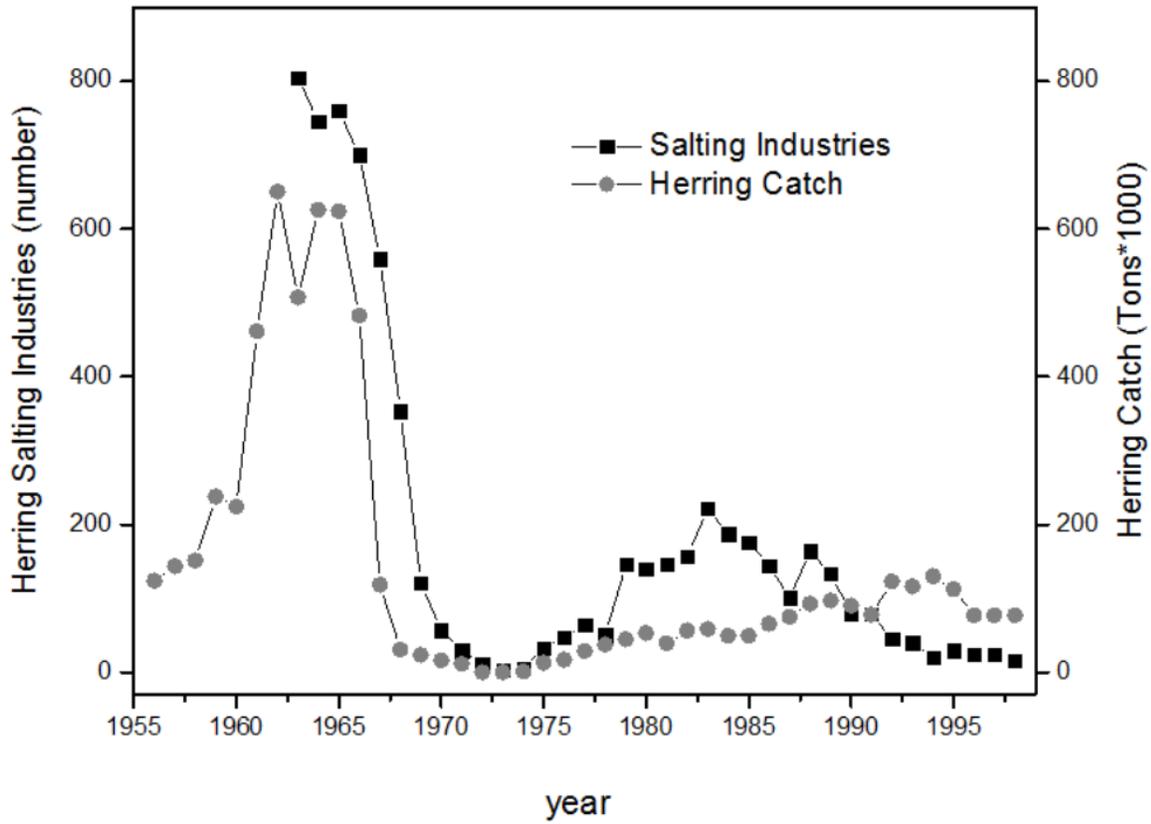

**Figure 6. Total Icelandic Herrings catch from 1956 to 1998**. Catch (in Tons*1000) in comparison to the number of Salting Herrings Industries. Data Source Statistics Iceland (http://www.statice.is/).

As response to the depletion of such important stocks, in 1984 Iceland has established the individual quotas system for fishing vessel, ITQs, that allocates maximum catch on the base of past years' vessel's catch performance. In recent years the country has adopted TAC (total allowable catch) regulation and other management measures such as area restrictions, fishing gear restrictions to conserve vulnerable habitats. Nevertheless, the curve of total catch of the Iceland from 1905 up to 2014, shows a peak of exploitation in late 80s (figure 7) followed by the capital peak, this time represented in form of economic index, the volume index at constant price.

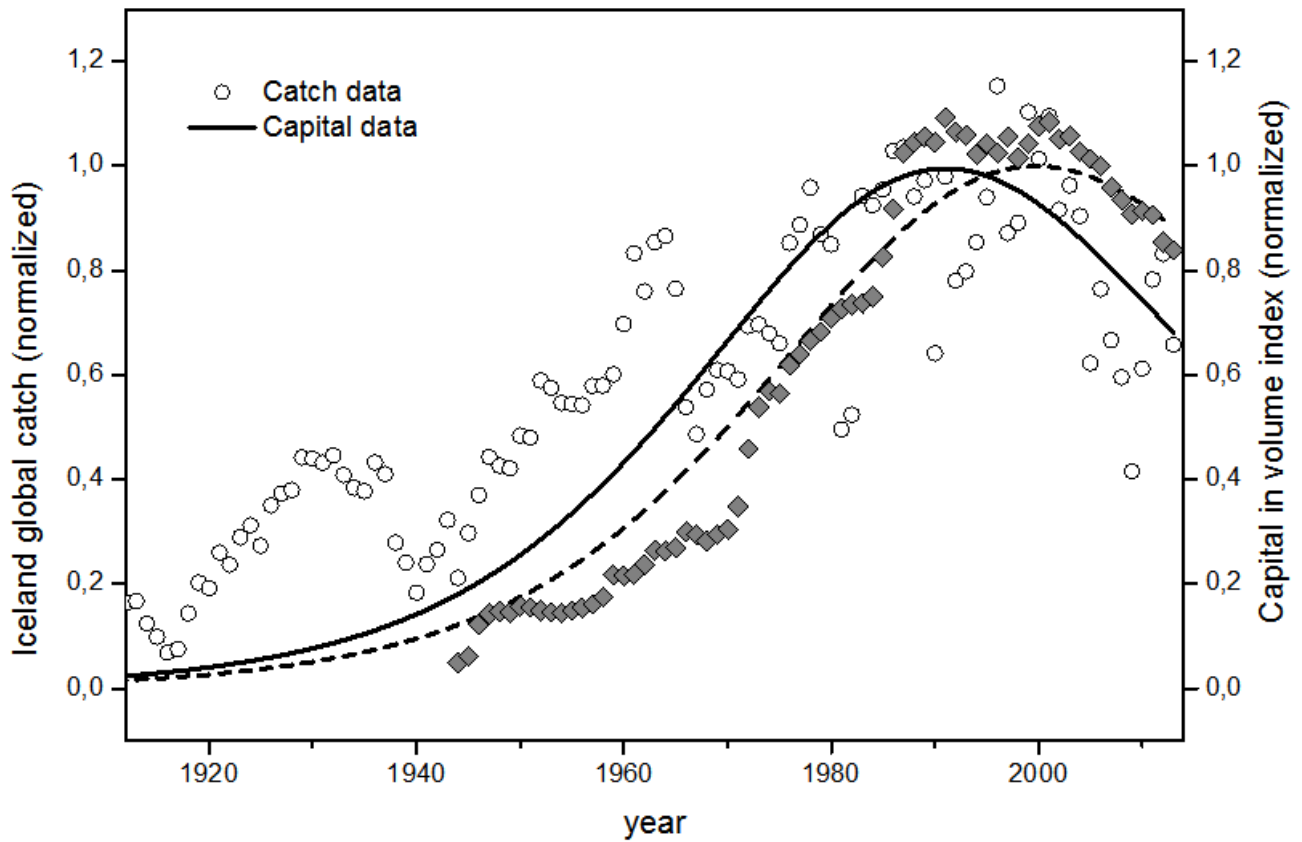

**Figure 7. Lotka Volterra modeling of Iceland Fishery Industry.** Total catch (production-prey) and Gross Fixed Capital Formation (GFCF) for Fishing expressed in Volume index (capital-predator) from 1905 to 2014. GFCF index data are available only from 1945. Normalizing factors: catch 1.64 $10^6$ Tons, Volume index at constant price 9.76 $10^1$ Million ISK/index (data 1990-2014 index=100 reference year 2005; data 1945-1990 index=100 reference year 1990). Data source: Statistics for Iceland (http://www.statice.is/). The GOF for the fitted data: NMSE Catch fit: 0.61; NMSE Index fit: 0.96.

**Other cases of overfishing.** There are many more cases in the literature where the behavior of the fishery can be interpreted in this way, although not always the typical bell shaped curve can be detected. Nevertheless, the evidence of overexploitation is still detectable in the fact that the "capital" parameter keeps growing for a period, while the production of the fishery declines. A good example is the that of the UK trawling fishery that saw a rapid decline after an intensive use of trawlers over 100 years[28]. The decline occurred despite the considerable increase of the "fishing power," something that can be likened to the parameter called here fishing capital (figure 8).

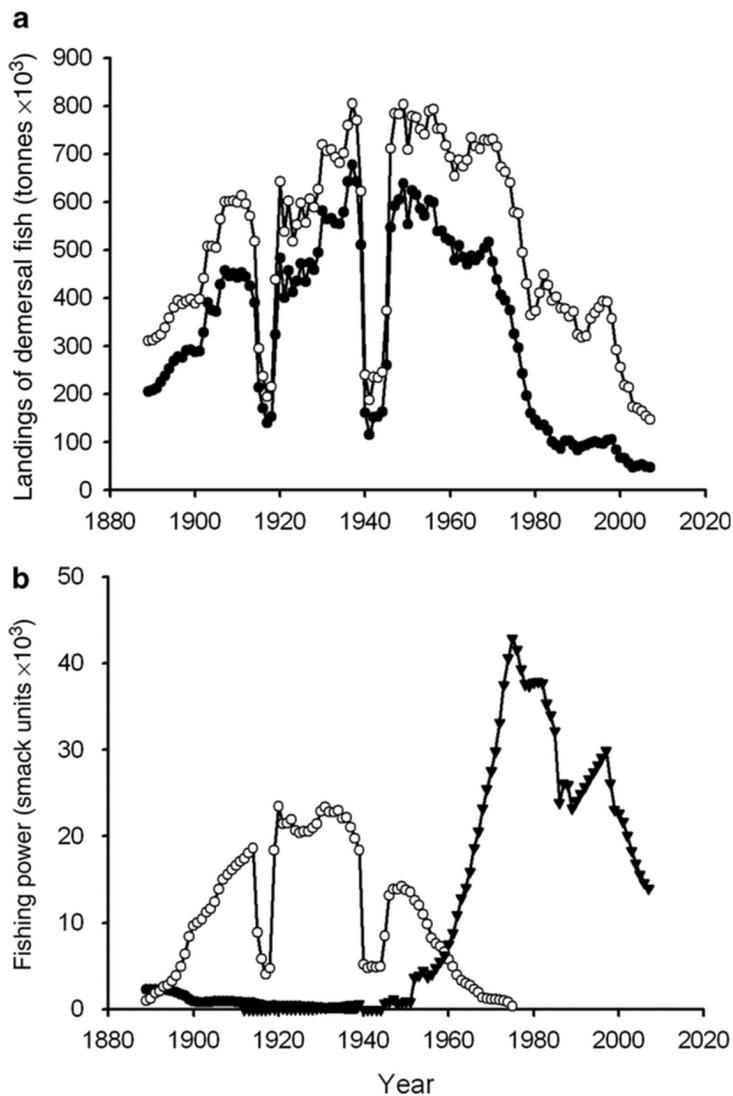

**Figure 8. The collapse of the UK trawling fishery.** By Thurstan et al., 2010. Note how the "fishing power" has continued to increase despite the decline of the catch. This is a typical feature of the model described in the present paper.

The scientific literature reports several other documents on historic fisheries collapses, as the famous case of North Atlantic cod collapse in Newfoundland (Hutchings & Myers, 1994)[29]. Then, the same behavior can be seen at the planetary level, where the LV model appears to be able to qualitatively describe the overexploitation. Statistics FAO (FishStat) shows that world catches reached the peak in the middle of the 1990s, then started declining; an ongoing phenomenon. Such a behavior has been correlated with the engine power of global fishing fleet (figure 9) [30], that is another "proxy" variable to measure the capital invested in fishing at World level. The work of Pauly&Zeller [31] also shows a similar behavior.

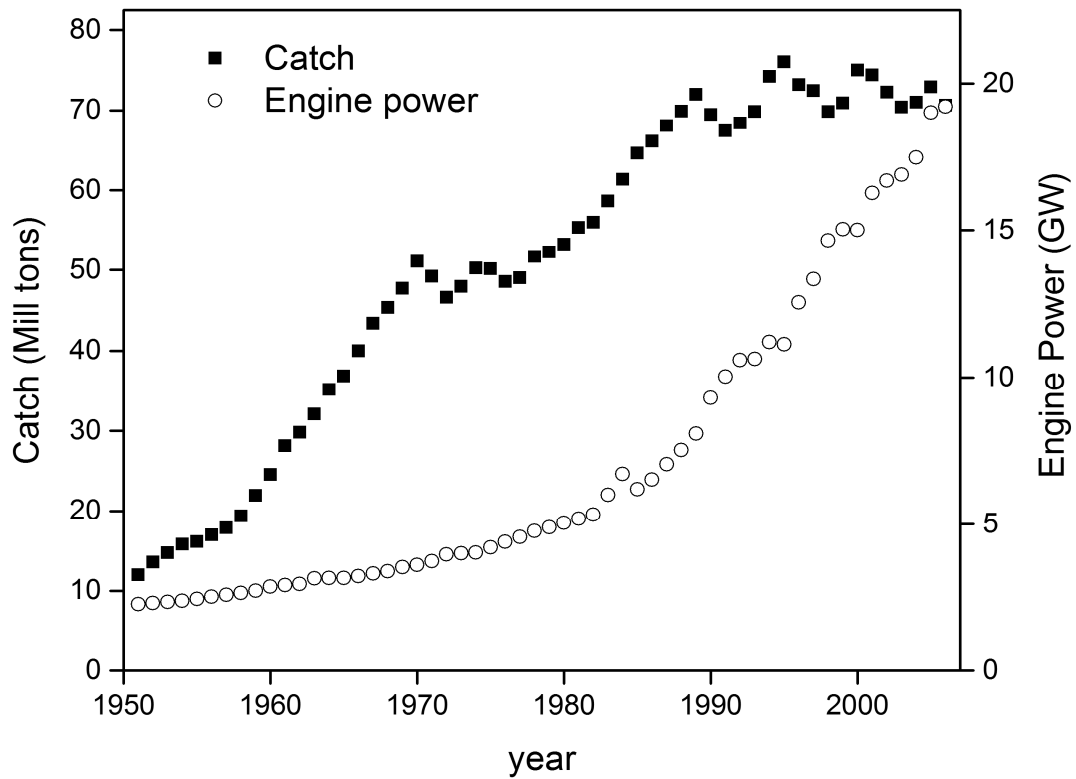

**Figure 9. World Fish Catch**. Trend in catch (Millions Tons) per unit of effort on power use (in Giga Watt) by fishing vessels (Pauly & Zeller, 2016).

**Discussion.**

The problem of overfishing is deeply debated by the scientific community and by market specialists., but often the problem the problem of overexploitation is underestimated or even ignored in the analysis, a tradition that may have started with the early study by Starbuck of the American whale fishery, where the decline of the catch was explained by such consideration as "the whales have become shy." However, from the analysis performed here, it appears that overexploitation is, indeed, a very important element in determining the behavior of fisheries. The dynamical model we used shows that, in many cases, the fishery yield is directly linked to the magnitude of the fish stock and that fishing continues all the way to bringing the yield to nearly zero by increasing the fishing effort despite fish depletion and consistently increased the fishing effort disregarding the risk of the collapse of the fish stock. These results evidence that today, despite the necessary efforts already put in the stock preservation, as fishing policy and regulations adopted by the majority of the countries, the fish demand on the market still drives the fishing pressure where fishing remains intensively practiced worldwide, with an increasing fishing power in terms of more sophisticated and expensive fishing techniques or equipment.

In the history of fisheries, the most common measure adopted to avoid the collapse of the fish stock has consisted in a total moratorium on fishing. In some cases, this approach has led to the stock 'salvation' as in the case of the Californian sardine (end of 1950), whaling (1985), the Canadian North Atlantic cod (1992), but it was a 'remedy' not a solution able to prevent the related fishery industry collapse. Moreover, the time lapse necessary for the recovery of stocks has often been quite long in comparison the time-scale of development of the modern fishing economy, that requires meeting the world demand. On this point it is sufficient to remember that the "right whale" fishery has not yet recovered from the intensive overexploitation it suffered in the 19$^{th}$ century[32]. Today, in order to prevent the collapse of fisheries, many international, national, and governmental organizations establish quotas for commercial stocks, expressed in terms of "total allowable catch" (TAC). This limit is deliberated on the basis of advice from scientific advisory bodies. Despite quotas, however, there exist several historical and recent cases where stock collapses still occurred, as shown by the case of the closure of the sardine fishing season off the length of the U.S. West Coast, recommended by the Pacific Fishery Management Council[33] last April (2016). It was closed mid-season in 2015 due to low stocks, but it has since fallen further. Federal rules mandate that the harvest must be closed if adult stocks fall below 150,000 tonnes, and the government estimates that there are now less than 65,000 tonnes.

The innovative approach of the present study is that it describes the quantitative correspondence between a biological resource and a capital by which the dynamic of overfishing can simply be represented. We are not just fitting data, but interpreting them with a model that specifies the dynamic interactions within its internal elements, so that obtaining a good fit indicates that the corresponding elements in the real world do behave according to the assumptions made in the model. Moreover, in the case of fishing, the model demonstrates how fisheries are set to perform their activities pointing a specific target, the fish species (whales, cod, sardine) thus the reproduction of predator is really dependent only on the abundance of those. The uniqueness is also evident even when we perform the analysis of fisheries at country level (UK, Japan, Iceland) in which the result curves of production and capital trends are due to a superimposition of the effects of every single fishery acting in that specific portion of fishing area. This uniqueness is not trivial in other world system, as demonstrated in the already cited case of the 'hare and lynx'. Michael Gilpin reported[34] the prey-predator relationship subsisted between these two populations until Canadian trappers entered in the dynamic to catch the lynx; after that, the data showed the hare was behaving as the predators and the lynx as the prey, reducing the validity of the model to represent a real hare-lynx dynamic but in absence of trappers.

We believe that our results are important in highlighting the fundamental role that overexploitation is playing in the cycle of modern fisheries, confirming the interpretation, for instance, by Daniel Pauly[35] and Pauly & Zeller[32] on global tendency. Overexploitation is a major problem that plagues the world's economy even for resources that are theoretically renewable, as highlighted, for instance, in the series of studies that were published under the title of "The Limits to Growth"[36]. This phenomenon needs to be understood if it is to be avoided by means of appropriate measures such as quotas, restrictions, or sustaining fisheries toward the transition to a circular economy.

**Conclusion**

The scope of the present study was to evidence how the abstract concept of "overexploitation" can be detected and analyzed by a dynamic model derived from the well know Lotka-Volterra, "prey/predator" model. The history of fisheries tells us of how the yield of the system is determined as the result of an intensive 'fish extraction' effort. Fishes are a renewable resource, and this is undoubtedly true, as long as the velocity of the resource reproduction is faster respect to the velocity of depletion. But in fact, in particular after the II world war, due to the rapid growth of the economy in an open access market, the 'velocity of fishing' (power fishing) in several different kind of fish supply chains, experienced a fast speed up in comparison to the 'velocity of the fish stock rebuilding' that remained almost invariant or even lowered, because more and more younger spawns were caught or trapped to satisfy the growing fish demand. Thus, in such a situation, a net resource outflow depletes the fish stock and this flow can not be balanced by the inflow due to the biological renewability of the resource itself. Here we show that the prey-predator dynamic can capture the behavior of the whole fishing sector of a country operating in a global open market framework. The innovative approach of this model is that it is a quantitative application of system dynamics that emphasizes the role of depletion and the feedback relationships of the various parameters of the models. The reasonably good fitting of the model with the historical data does not mean that depletion is the only forcing that affects the system, but it invites to consider depletion as an important parameter even in systems that are normally defined as "renewable".

**Acknowledgments.** This work was partially supported by the MEDEAS project, funded by the European Union's Horizon 2020 research and innovation program under grant agreement No 691287. The opinion expressed in the present work are those of the authors' only and are not to be attributed to any organ of the European Union.**References.**

1.  Lotze, H. K. & Worm, B. Historical baselines for large marine animals. Trends Ecol. Evol. 24, 254–262 (2009), http://doi.org/10.1016/j.tree.2008.12.004

2.  Bailey, J. Adventures in cross-disciplinary studies: Grand strategy and fisheries management, *Mar. Policy*, **63**, 18–27 (2016).

3.  Gordon, H. S. The economic theory of a common-property resource: the fishery. *J. Polit. Econ.* **62**, 124–142 (1954).

4.  Schaefer, M. B. Some Considerations of Population Dynamics and Economics in Relation to the Management of the Commercial Marine Fisheries. *J. Fish. Res. Board Can.* **14**, 669–681 (1957).

5.  Hardin, G. The tragedy of the commons. *Science* **162**, 1243–1248 (1968).

6.  Roopnarine, P. Ecology and the Tragedy of the Commons. *Sustainability*, **5**, 349–773 (2013).

7.  Schoener, T. W. Alternatives to Lotka-Volterra Competition: Models of Intermediate Complexity. *Theor. Popul. Bio.* **10**, 309–333. (1976).

8.  Worm, B. et al. Rebuilding global fisheries. *Science,* **325**, 578–85 (2009), http://doi.org/10.1126/science.1173146

9.  Perry L., Low P.J., Ellis J.R., Reynolds J.D. Climate Change and Distribution Shifts in Marine Fishes , Science, published online 12 May 2005, DOI: 10.1126/science.111132220